\pdfoutput=1
\documentclass[twocolappendix]{emulateapj}
\usepackage{amsmath}
\usepackage{graphicx,epstopdf}
\usepackage{multirow}
\usepackage{color}
\usepackage{latexsym}
\usepackage{amssymb}
\usepackage{epsfig}
\usepackage{verbatim}
\usepackage{placeins}

\usepackage[colorlinks=true,linkcolor=blue,citecolor=blue]{hyperref}

\begin{document}

\title{Super-Earth Atmospheres: Self-Consistent Gas Accretion and Retention}

\author{Sivan Ginzburg\altaffilmark{1}, Hilke E. Schlichting\altaffilmark{2,3,4}, Re'em Sari\altaffilmark{1}}

\altaffiltext{1}{Racah Institute of Physics, The Hebrew University, Jerusalem 91904, Israel}

\altaffiltext{2}{Massachusetts Institute of Technology, 77 Massachusetts Avenue, Cambridge, MA 02139-4307, USA}

\altaffiltext{3} {UCLA, 595 Charles E. Young Drive East, Los Angeles, CA 90095, USA}

\altaffiltext{4} {California Institute of Technology, 1200 E. California Blvd., Pasadena, CA 91125, USA}

\begin{abstract}
Some recently discovered short-period Earth to Neptune sized exoplanets (super Earths) have low observed mean densities which can only be explained by voluminous gaseous atmospheres. Here, we study the conditions allowing the accretion and retention of such atmospheres. We self-consistently couple the nebular gas accretion onto rocky cores and the subsequent evolution of gas envelopes following the dispersal of the protoplanetary disk. Specifically, we address mass-loss due to both photo-evaporation and cooling of the planet. We find that planets shed their outer layers (dozens of percents in mass) following the disk's dispersal (even without photo-evaporation), and their atmospheres shrink in a few Myr to a thickness comparable to the radius of the underlying rocky core. 
At this stage, atmospheres containing less particles than the core (equivalently, lighter than a few \% of the planet's mass) can be blown away by heat coming from the cooling core, while heavier atmospheres cool and contract on a timescale of Gyr at most.
By relating the mass-loss timescale to the accretion time, we analytically identify a Goldilocks region in the mass-temperature plane in which low-density super Earths can be found: planets have to be massive and cold enough to accrete and retain their atmospheres, while not too massive or cold, such that they do not enter runaway accretion and become gas giants (Jupiters). We compare our results to the observed super-Earth population and find that low-density planets are indeed concentrated in the theoretically allowed region. Our analytical and intuitive model can be used to investigate possible super-Earth formation scenarios.
\end{abstract}


\section{Introduction}\label{sec:introduction}

In recent years, the {\it Kepler} mission has discovered many Earth to Neptune size exoplanets \citep[see, e.g.,][]{WeissMarcy2014}. Some of these ``super Earths'', or ``mini Neptunes'', have low densities which rule out a purely rocky composition. The low density can be explained by either a gaseous Hydrogen/Helium envelope atop a rocky core, or by a water rich composition. However, at least some super Earths have densities low enough, that a gaseous atmosphere is essential to explain their inferred masses and radii \citep{Lopez2012,Lissauer2013}.

The existence of significant gaseous envelopes surrounding rocky cores fits in naturally in the context of core-nucleated accretion theory. According to this theory, gas giants such as Jupiter formed by gas accretion onto solid cores from the gas-rich nebula which surrounded the young sun \citep{PerriCameron1974,Harris1978,Mizuno1978,Mizuno1980,Stevenson1982}. The rate by which a rocky core accumulates gas is determined by the atmosphere's cooling rate (Kelvin-Helmholtz contraction), until the acquired atmosphere is comparable in mass to the core, at which stage a runaway accretion initiates and the planet explodes into a gas giant \citep{BodenheimerPollack86,Pollack96,Lee2014,PisoYoudin2014,LeeChiang2015,Piso2015}. The amount of gas a rocky core embedded in the protoplanetary gas nebula acquires (and whether it will explode into a gas giant) is therefore determined by comparing the atmosphere's cooling timescale to the gas disk lifetime \citep[a few Myr; see, e.g.,][]{Mamajek2009,WilliamsCieza2011,Alexander2014}.
In some cases, however, the assembly of the rocky core cannot be decoupled from the accretion of gas, and the effects of planetesimal impacts \citep{Pollack96,Rafikov2006,Rafikov2011}, or even giant collisions of protoplanets \citep{InamdarHilke}, have to be taken into account, depending on the orbital separation from the parent star.

In this work we focus on super Earths in close orbits of $\sim 0.1$ AU, with the Kepler-11 system \citep{Lissauer2011} as a well-studied example. At least some of these planets require Hydrogen/Helium envelopes which constitute $\sim$ 1\% to 10\% of the planet's mass to explain the observations \citep{Lopez2012}. While some works argue that such heavy atmospheres are difficult to accrete in-situ during the gas disk's lifetime \citep{InamdarHilke}, other works \citep{Lee2014} raise the opposite concern, that accretion is too efficient, making super Earths vulnerable to runaway and becoming gas giants. Nonetheless, by examining the atmosphere mass growth with time in both \citet{Lee2014} and \citet{InamdarHilke}, we find that rocky cores of a few $M_{\Earth}$ acquire gas envelopes of a few \% during a disk lifetime of a few Myr, at least for some plausible nebula parameters, and disregarding the effects of giant impacts (perhaps due to inward migration of an assembled core during the disk lifetime, or a solid-enhanced nebula). Therefore, low density planets with such masses are a natural ``missing link'' between smaller rocky planets, which did not manage to accrete a substantial atmosphere, and more massive cores \citep[a fiducial critical value of $\approx 10M_{\Earth}$ is often cited; see, e.g.,][]{Rafikov2006,Rafikov2011,Lee2014}, which gathered their own mass in gas, and exploded into gas giants.

Although super Earths with gas envelopes of a few \% are a plausible outcome of nebular gas accretion, their evolution continues after the disk dispersal. In particular, many studies have demonstrated that these gas envelopes are susceptible to significant evaporation and mass loss due to high-energy stellar photons once the nebula has dispersed \citep{Lopez2012,OwenJackson2012,OwenWu2013}.

In this work we emphasize the role of the planet's cooling luminosity as an additional energy source for driving mass loss \citep[see also][]{IkomaHori2012,OwenWu2015}, and relate the post-dispersal evolution to the preceding accretion, in order to obtain constraints on the possible formation scenarios of these planets. By self-consistently coupling the nebular accretion to the subsequent mass-loss once the nebula has dispersed \citep[see also][]{Rogers2011}, we find limiting conditions in which super Earths can acquire and hold on to their atmospheres.  

The outline of the paper is as follows. In Section \ref{sec:accretion} we discuss the accretion of gas from the nebula. In Section \ref{sec:dispersal} we study the subsequent evolution once the gas nebula has dispersed. Section \ref{sec:photo_evap} is devoted to atmosphere evaporation by high-energy photons, and to how it sculpts the super-Earth population. Section \ref{sec:obs} compares our theoretical constraints to the observations, and our conclusions are summarized in Section \ref{sec:conclusions}.

\section{Nebular Gas Accretion}\label{sec:accretion}

Gas accretion by rocky cores has been studied extensively, both numerically and analytically \citep{BodenheimerPollack86,Pollack96,BodenheimerLissauer2014,Lee2014,PisoYoudin2014,LeeChiang2015,Piso2015}.
Here, we briefly summarize the main concepts outlined in these works, and provide a simplified analytical description of nebular gas accretion. Since our main goal is to provide intuition for the subsequent phases of planetary evolution, we adopt many simplifying assumptions in comparison with previous studies. 

The formation timescale of a rocky protoplanet is much shorter than the gas nebula lifetime in the inner disk \citep{GLS,Lee2014}, leading us to ignore the impact of planetesimals during gas accretion. However, the isolation mass of a protoplanet is also small in the inner disk \citep{GLS,InamdarHilke}, leading \citet{InamdarHilke} to consider giant impacts of protoplanets, once the gas has begun to disperse. Here, we ignore giant impacts and assume that gas accumulates onto an assembled rocky core (which may have migrated from larger distances, or created in a solid-enhanced nebula).

Throughout the paper, we assume that the self gravity of the atmosphere is negligible. Naively, when this condition breaks down runaway accretion occurs, leading to the formation of gas giants, which are not the focus of this work. More concretely, by inspecting Figure 2 of \citet{PisoYoudin2014}, we find that self-gravity affects the results by more than a factor of 2 for atmosphere mass fractions as low as 20\%, which are at the high end of the gas envelopes that we consider. For these atmospheres, our approximation is marginal.

\subsection{Adiabatic Atmosphere}\label{subsec:adiabatic}

The initial atmosphere a core of mass $M_c$ and radius $R_c$ accretes, in a short (dynamical) timescale, is given by an adiabatic profile, which did not have time to cool, and has the same entropy as the surrounding nebula. Hydrostatic equilibrium sets the following adiabatic temperature profile \citep[see also][]{Rafikov2006,PisoYoudin2014,InamdarHilke}
\begin{equation}\label{eq:t_adiabat}
\frac{T(r)}{T_{\rm eq}}=1+\frac{R_{\rm B}'}{r}-\frac{R_{\rm B}'}{r_{\rm out}}
\approx
\begin{cases}
1 & r=r_{\rm out}\\
R_{\rm B}'/r_{\rm out} & r\lesssim r_{\rm out} \\
R_{\rm B}'/r & r\ll r_{\rm out}
\end{cases},
\end{equation}
where $T_{\rm eq}\sim10^3\textrm{ K}$ is the temperature of the surrounding nebula, and $R_{\rm B}'=(1-1/\gamma) R_{\rm B}\sim R_{\rm B}$, with $R_{\rm B}\simeq GM_c\mu/k_{\rm B}T_{\rm eq}$ denoting the Bondi radius. $G$ is the gravitation constant, $k_{\rm B}$ is Boltzmann's constant, $\mu$ the molecular mass, and $\gamma$ the adiabatic index, which we choose as $\gamma=7/5$, suitable for diatomic gas. \citet{Lee2014} and \citet{Piso2015} incorporate more elaborate equations of state into their numerical calculations, in which $\gamma$ varies with temperature and pressure. Nonetheless, we approximate here $\gamma$ as constant for simplicity and discuss the sensitivity of the results to this choice below. $r_{\rm out}=\min(R_{\rm B},R_{\rm H})$ denotes the radius where the planet's atmosphere blends in with the surrounding nebula, which is taken to be the minimum of the Bondi radius and the Hill radius $R_{\rm H}=a(M_c/M_\Sun)^{1/3}$, with $a$ denoting the semi-major axis. For a few $M_\Earth$ core at $\sim 0.1$ AU, $R_{\rm B}\approx R_{\rm H}\sim 10^{10}\textrm{ cm}$, with less massive or more distant cores having $R_{\rm B}<R_{\rm H}$ \citep[see, e.g.,][]{Rafikov2006,InamdarHilke}.
Note that the nebular scale height is $z_d=c_s/\Omega\gtrsim 10^{10}\textrm{ cm}$, with $c_s\simeq(k_{\rm B}T_{\rm eq}/\mu)^{1/2}$ denoting the sound speed and $\Omega=(GM_\Sun/a^3)^{1/2}$ the orbital period \citep[see, e.g.,][]{Hayashi1981}. Therefore we may assume spherical accretion, although this assumption is marginal.
The adiabatic density profile is
\begin{equation}\label{eq:rho_adiabat}
\frac{\rho(r)}{\rho_d}=\left(1+\frac{R_{\rm B}'}{r}-\frac{R_{\rm B}'}{r_{\rm out}}\right)^{1/(\gamma-1)},
\end{equation}
with $\rho_d=\sigma_d/z_d\sim 10^{-6}\textrm{ g cm}^{-3}$ denoting the disk density, assuming a minimum mass solar nebula (MMSN) surface density $\sigma_d\approx 5\cdot 10^4(T_{\rm eq}/10^3\textrm{ K})^3\textrm{ g cm}^{-2}$ \citep{Hayashi1981}, although the {\it Kepler} planets might have formed in a more massive or solid-enhanced nebula \citep{ChiangLaughlin2013,Hilke2014}.

By integrating Equation \eqref{eq:rho_adiabat} we find that for $\gamma>4/3$ the mass fraction of the atmosphere is given by
\begin{equation}\label{eq:mass_adiabatic}
\begin{split}
f&\equiv\frac{M_{\rm atm}}{M_c}\sim\frac{\rho_d r_{\rm out}^3}{M_c}\left(\frac{R_{\rm B}}{r_{\rm out}}\right)^{1/(\gamma-1)}
\\&\sim 10^{-3}\cdot
\begin{cases}
{\displaystyle
\left(\frac{M_c}{M_\Earth}\right)^2\left(\frac{T_{\rm eq}}{10^3{\textrm K}}\right)^{5/2}} & 
{\displaystyle R_{\rm B}<R_{\rm H}}\\
{\displaystyle \left(\frac{M_c}{M_\Earth}\right)^{5/3}\left(\frac{T_{\rm eq}}{10^3{\textrm K}}\right)^2} & {\displaystyle R_{\rm B}>R_{\rm H}}
\end{cases},
\end{split}
\end{equation}
where we emphasize the transition between Bondi and Hill boundary conditions, at a few $M_\Earth$, and where the factor $(R_{\rm B}/r_{\rm out})^{1/(\gamma-1)}$ is a correction to similar expressions by \citet{ChiangLaughlin2013} and \cite{InamdarHilke} for $r_{\rm out}=R_{\rm H}$, due to the density jump at $r\approx r_{\rm out}$, evident in Equation \eqref{eq:rho_adiabat}. Equation \eqref{eq:mass_adiabatic} shows that adiabatic gas accretion is insufficient and cannot explain observed envelope mass fractions without invoking a massive nebula \citep[by 1-2 orders of magnitude; see, e.g.,][]{ChiangLaughlin2013,BodenheimerLissauer2014, InamdarHilke}. Equation \eqref{eq:mass_adiabatic} also indicates that the assembly of $N$ small envelopes (especially in the $r_{\rm out}=R_{\rm B}$ regime) into a large one during the giant collisions phase is inefficient (as $N^{-2}$), even if we ignore mass loss due to collisions. 

The gas fraction for $r_{\rm out}=R_{\rm H}\sim R_{\rm B}$ (relevant for a few $M_\Earth$ cores at $\sim 0.1$ AU) can be written as \citep[see][]{ChiangLaughlin2013}
\begin{equation}\label{eq:max_f_adiabatic}
f\sim\frac{\rho_d a^3}{M_\Sun}\sim\frac{G\sigma_d}{c_s\Omega}\sim Q^{-1},
\end{equation}
where $Q>1$ is Toomre's stability criterion for the gas disk. \citet{ChiangLaughlin2013} disregard the $(R_{\rm B}/r_{\rm out})^{1/(\gamma-1)}$ factor in Equation \eqref{eq:mass_adiabatic}, leading to a constant $f$ as a function of $M_c$ in the $R_{\rm  B}>R_{\rm H}$ regime, and to their claim that Equation \eqref{eq:max_f_adiabatic} represents the maximal gas fraction, in contrast to Equation \eqref{eq:mass_adiabatic} which predicts a further increase in $f$ for larger core masses ($R_{\rm B}>R_{\rm H}$).  

Equation \eqref{eq:max_f_adiabatic} demonstrates that cores a few times the mass of Earth can accrete arbitrarily large (up to runaway) adiabatic atmospheres, if embedded in a massive enough nebula \citep[as invoked by][]{ChiangLaughlin2013,Hilke2014}. Nonetheless, we assume a light (MMSN, or more precisely, its extrapolation beyond Mercury's orbit) gas nebula, in which adiabatic atmospheres are small, as indicated by Equation \eqref{eq:mass_adiabatic}, and further accretion by the cooling of the envelope has to be considered.  In order to reconcile this gas-poor nebula with the {\it Kepler} observations (specifically, the high solid surface densities required to assemble large cores at close distances), it might be necessary to consider either a large solid/gas ratio or inward migration of rocky cores \citep{Hilke2014}.

\subsection{Accretion by Cooling}\label{subsec:accretion_cooling}

\begin{figure}[tbhp]
	\includegraphics[width=\columnwidth]{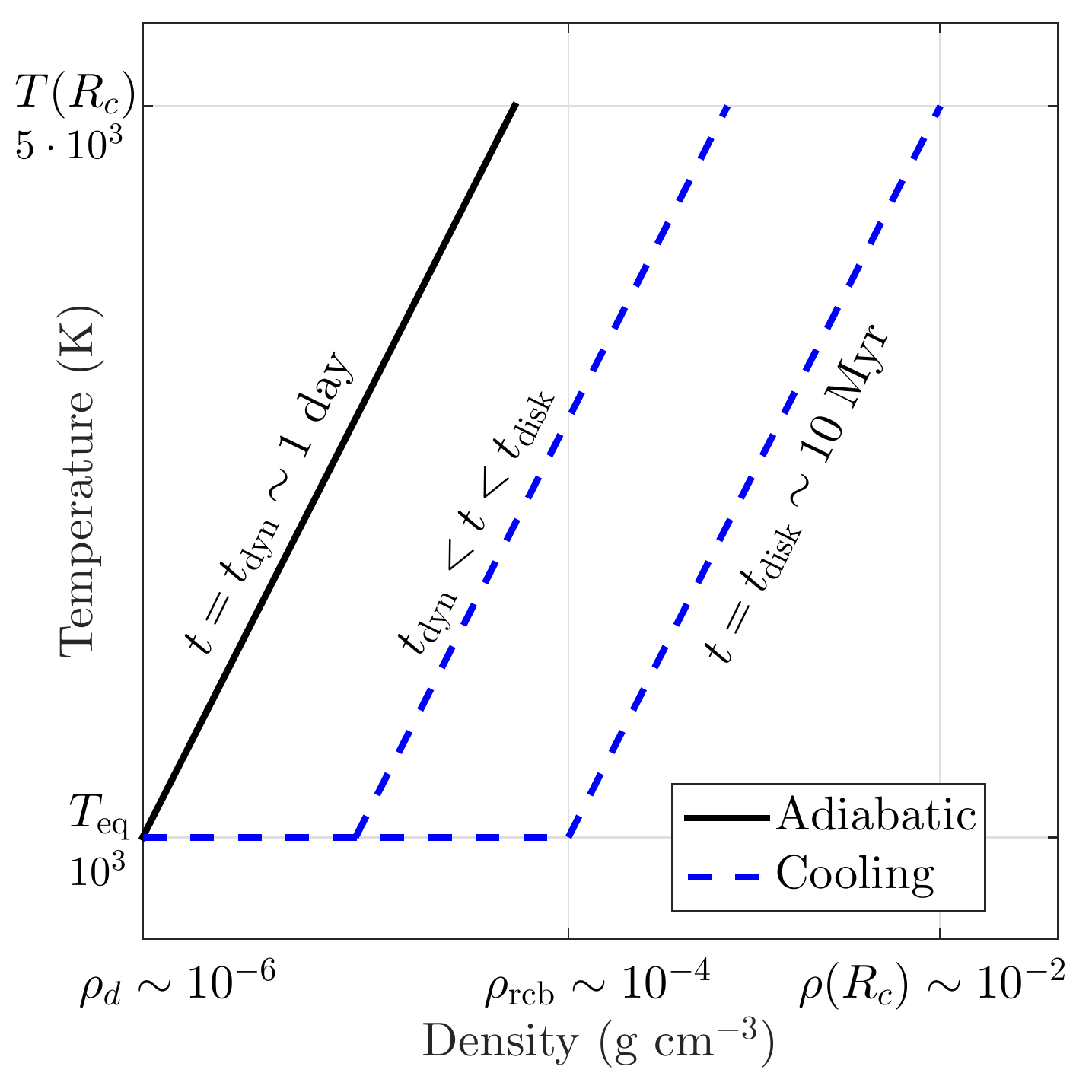}
	\caption{Schematic temperature vs. density profiles (logarithmic scale) of a super-Earth atmosphere during the nebular accretion phase. The initial adiabatic atmosphere (solid black line) is isentropic, while at later stages (2 successive profiles are plotted) the cooling (and accreting) envelope is characterized by a nearly isothermal radiative outer layer, and a convective interior (dashed blue lines). Typical values of the density and temperature are provided.
		\label{fig:scheme_accretion}}
\end{figure}

As the gas envelope cools through radiation, an outer radiative layer develops, while the interior remains isentropic, due to convection \citep[see, e.g.,][]{Lee2014,PisoYoudin2014,LeeChiang2015}. The convective profile is similar to Equation \eqref{eq:t_adiabat}
\begin{equation}\label{eq:t_convective}
\frac{T(r)}{T_{\rm rcb}}=1+\frac{R_{\rm B}'}{r}-\frac{R_{\rm B}'}{R_{\rm rcb}},
\end{equation}
with $T_{\rm rcb}$, $R_{\rm rcb}$ denoting the temperature and radius of the radiative-convective boundary (RCB), respectively, and with $R_{\rm B}'$ now denoting (we use the same notation, since $T_{\rm rcb}\sim T_{\rm eq}$, up to an order of unity factor which we disregard, as we show below)
\begin{equation}\label{eq:bondi}
R_{\rm B}'\equiv\frac{\gamma-1}{\gamma}\frac{GM_c\mu}{k_{\rm B}T_{\rm rcb}}.
\end{equation}
Similarly to Equation \eqref{eq:rho_adiabat}, the density profile is given by
\begin{equation}\label{eq:rho_convective}
\frac{\rho(r)}{\rho_{\rm rcb}}=\left(1+\frac{R_{\rm B}'}{r}-\frac{R_{\rm B}'}{R_{\rm rcb}}\right)^{1/(\gamma-1)},
\end{equation}
with $\rho_{\rm rcb}$ marking the density at the RCB.
\citet{Lee2014} find that the RCB is determined by ${\rm H}_2$ dissociation, at an almost fixed temperature of 2500 K. Here, we adopt a different approach, and assume power law opacities \citep[similar to][]{Rafikov2006,PisoYoudin2014}, which are relevant for dust-free models \citep{LeeChiang2015}, and lead to $T_{\rm rcb}\sim T_{\rm eq}$, up to an order of unity coefficient, which we omit in our approximate analysis. For $r\ll R_{\rm rcb}<R_{\rm B}'$ (we make a further approximation, and assume that the outer boundary condition is always dictated by $R_{\rm B}$)
\begin{equation}
k_{\rm B}T(r)\approx\frac{\gamma-1}{\gamma}\frac{GM_c\mu}{r},
\end{equation}
or intuitively, the scale height is $h\sim r$ (leading to a  power law density profile),
and specifically, the temperature at the atmosphere-core boundary $T(R_c)$ remains constant 
at $k_{\rm B}T(R_c)\sim GM_c\mu/R_c$, with $R_c\sim 10^9\textrm{ cm}\ll R_{\rm B}$. At this temperature, the rocky core is molten, and is therefore convective. Since the core is convective (i.e. has a uniform entropy) and almost incompressible, we approximate it as isothermal. This simplification is in accordance with realistic adiabatic profiles, in which the temperature changes only by a factor of order unity, while the pressure varies by orders of magnitude \citep[e.g.,][]{Katsura2010}. Because $T(R_c)$ is constant, as explained above, core cooling plays no role in the energy budget as long as $R_c\ll R_{\rm rcb}$.
The specific energy in the interior is therefore
\begin{equation}\label{eq:de_dm}
e=-\frac{GM_c}{r}+\frac{1}{\gamma-1}\frac{k_{\rm B}T}{\mu}=-\frac{\gamma-1}{\gamma}\frac{GM_c}{r},
\end{equation}
implying that for $\gamma<3/2$ the total energy is concentrated in the inside, and is given by
\begin{equation}\label{eq:enr_atm}
E=-\frac{\gamma-1}{\gamma}\frac{\gamma-1}{3-2\gamma}4\pi R_c^2GM_c\rho(R_c).
\end{equation}
According to Equation \eqref{eq:enr_atm}, as the envelope cools, its density increases, and therefore the radiative (and nearly isothermal) region thickens, as depicted schematically in Figure \ref{fig:scheme_accretion}.

The total envelope mass, on the other hand, is concentrated in the outside of the convective region, $r\sim R_{\rm rcb}$, for $\gamma>4/3$. By integrating Equation \eqref{eq:rho_convective} we find
\begin{equation}\label{eq:mass_atm}
M_{\rm atm}=A(\gamma)4\pi R_{\rm rcb}^3\rho_{\rm rcb}\left(\frac{R_{\rm B}'}{R_{\rm rcb}}\right)^{1/(\gamma-1)},
\end{equation}
where the numerical coefficient $A(\gamma=7/5)=5\pi/16$. The factor of $(R_{\rm B}'/R_{\rm rcb})^{1/(\gamma-1)}$ is due to the temperature and density jump at $r\approx R_{\rm rcb}$ (where the scale height is $R_{\rm rcb}^2/R_{\rm B}<R_{\rm rcb}$). The mass in the radiative (nearly isothermal) region is negligible due to the exponentially decreasing density \citep[see also][]{PisoYoudin2014}. \citet{LeeChiang2015} argue for $\gamma<4/3$, due to ${\rm H}_2$ dissociation \citep[see also][]{Piso2015}, and based on numerical calculations in \citet{Lee2014}. For such values of $\gamma$, the mass is also concentrated in the inside (as the energy), and some of our results change. Nonetheless, we ignore Hydrogen dissociation here for simplicity, and present results for the diatomic $\gamma=7/5$ \citep[see][for a similar model]{PisoYoudin2014}. Qualitative conclusions which are sensitive to our choice of $\gamma>4/3$ are discussed below, and the alternative solution for $\gamma<4/3$ is also provided. We relate $\rho(R_c)$ to $\rho_{\rm rcb}$ adiabatically (see Figure \ref{fig:scheme_accretion}), and combine Equations \eqref{eq:enr_atm} and \eqref{eq:mass_atm}
\begin{equation}\label{eq:energy_mass_radius}
E=-\frac{(\gamma-1)^2}{A(\gamma)\gamma(3-2\gamma)}\frac{GM_cM_{\rm atm}}{R_c}\left(\frac{R_{\rm rcb}}{R_c}\right)^{-(3\gamma-4)/(\gamma-1)}.
\end{equation}
Equation \eqref{eq:energy_mass_radius} demonstrates that cooling of the envelope corresponds either to mass increase \citep[``to cool is to accrete'', as phrased by][]{LeeChiang2015} or to radius decrease. As we show below, up to logarithmic factors, cooling is indeed equivalent to gas accretion.

We relate the radius $R_{\rm rcb}$ to the RCB density $\rho_{\rm rcb}$ using hydrostatic equilibrium of the radiative region between $R_{\rm rcb}$ and $R_{\rm B}$ \citep[see also][]{PisoYoudin2014}
\begin{equation}\label{eq:radius_rho_d}
\frac{R_{\rm rcb}}{R_{\rm B}}=\frac{1}{1+\ln{(\rho_{\rm rcb}/\rho_d)}}.
\end{equation} 
Equation \eqref{eq:radius_rho_d} implies that the atmosphere can increase in density by orders of magnitude, while the radius shrinks only by a logarithmic factor. Therefore, the simplest approximation for the gas accretion phase is to assume that the radius remains roughly constant $R_{\rm rcb}\approx R_{\rm B}$. The atmosphere mass in this approximation, using Equation \eqref{eq:mass_atm}, is given by $M_{\rm atm}\approx\rho_{\rm rcb}R_{\rm B}^3$, and by comparison with Equation \eqref{eq:mass_adiabatic} we find that the gas fraction $f$ increases by a ratio of $\rho_{\rm rcb}/\rho_d$ relative to the adiabatic atmosphere. Since we are interested in atmospheres of a few \% in mass, and adiabatic atmospheres have $f\sim 10^{-3}$, we deduce that $\rho_{\rm rcb}/\rho_d\sim 10^1-10^2$, and $R_{\rm rcb}$ decreases (from an initial $R_{\rm B}$) by a factor of a few. With this approximation Equation \eqref{eq:energy_mass_radius} can be written as
\begin{equation}
E\sim-\frac{GM_cM_{\rm atm}}{R_c}\left(\frac{R_{\rm B}'}{R_c}\right)^{-(3\gamma-4)/(\gamma-1)},
\end{equation}
emphasizing that cooling is equivalent to gas accretion. This result differs from \citet{LeeChiang2015} by a factor of a few, given by $(R_{\rm B}'/R_c)^{1/2}$, due to their effective choice of $\gamma<4/3$.

The cooling rate (luminosity) of the envelope is given by combining the diffusion equation and hydrostatic equilibrium at the RCB
\begin{equation}\label{eq:lum}
L=\frac{64\pi}{3}\frac{\sigma T_{\rm rcb}^4R_{\rm B}'}{\kappa\rho_{\rm rcb}},
\end{equation}
with $\sigma$ denoting the Stephan-Boltzmann constant and $\kappa$ the opacity at the RCB. This result can be intuitively understood by $L\sim\sigma T_{\rm rcb}^4 R_{\rm rcb}^2/\tau$, with $\tau\sim\kappa\rho_{\rm rcb}h$ denoting the optical depth at the RCB, and $h=R_{\rm rcb}^2/R_{\rm B}$ the scale height there. We write an evolution equation $L=-\dot{E}$, and combine Equations \eqref{eq:mass_atm}, \eqref{eq:energy_mass_radius}, and \eqref{eq:lum}, to obtain the time it takes the envelope to reach a mass $M_{\rm atm}$
\begin{equation}\label{eq:time_mass}
t=B(\gamma)\frac{\kappa M_{\rm atm}^2}{\sigma R_c(GM_c)^3}\left(\frac{k_{\rm B}}{\mu}\right)^4\left(\frac{R_{\rm rcb}^2}{R_{\rm B}'R_c}\right)^{-(3\gamma-4)/(\gamma-1)},
\end{equation}
where the $\gamma$ dependent numerical factor is given by
\begin{equation}
B(\gamma)=\frac{3}{512\pi^2}\frac{\gamma^3}{(\gamma-1)^2(3-2\gamma)}A^{-2}(\gamma)=0.05,
\end{equation}
and evaluated for $\gamma=7/5$. By substituting $\kappa\approx 0.1\textrm{ cm}^2\textrm{ g}^{-1}$ \citep{Allard2001,Freedman2008}, and using the $R_{\rm rcb}\approx R_{\rm B}$ approximation, we rewrite Equation \eqref{eq:time_mass} and find the gas mass fraction at the time of disk dispersal $t_{\rm disk}$
\begin{equation}\label{eq:f_time}
f\approx 0.02\left(\frac{M_c}{M_\Earth}\right)^{0.8}\left(\frac{T_{\rm eq}}{10^3\textrm{ K}}\right)^{-0.25}\left(\frac{t_{\rm disk}}{1\textrm{ Myr}}\right)^{0.5},
\end{equation}
where we have ignored, for simplicity, the weak dependence of the opacity on density and temperature in the relevant regime, although it enables the transition into a convective region \citep[see][for a more elaborate discussion of opacities]{LeeChiang2015}. In the derivation of Equation \eqref{eq:f_time}, and throughout the paper, we used $R_c\propto M_c^{1/4}$ rather than the constant-density relation $R_c\propto M_c^{1/3}$ for the rocky core, in order to take into account the slight gravitational compression \citep[see, e.g.,][]{Valencia2006}, although both mass-radius relations yield a similar result due to the weak dependence on the core's radius $f\propto R_c^{1/4}$. 

Equation \eqref{eq:f_time} roughly reproduces the results of previous studies \citep{Lee2014,InamdarHilke,LeeChiang2015}. Specifically, we emphasize that the amount of accreted gas does not depend on the density of the nebula $\rho_d$, except for logarithmic factors due to equation \eqref{eq:radius_rho_d}, which we have neglected. Instead, the mass accretion is regulated by the envelope's own cooling time, while the radiative layer decouples the bulk envelope from the surrounding nebula's density boundary condition (see Figure \ref{fig:scheme_accretion}). This logarithmic dependence on the nebular density allows \citet{LeeChiang3} to accrete substantial gas envelopes even in highly depleted transitional disks. The power law $f\propto t^{1/2}$ \citep[see also][]{PisoYoudin2014,LeeChiang2015} is intuitively understood by noticing that while the energy is proportional to $M_{\rm atm}$, so is the optical depth, and therefore the luminosity is proportional to $M_{\rm atm}^{-1}$. Quantitatively, Equation \eqref{eq:f_time} predicts that super Earths will acquire atmospheres of a few \% in mass during the gas disk's lifetime, outweighing their initial adiabatic envelopes (which are proportional to $\rho_d$) in a MMSN (see Section \ref{subsec:adiabatic}). 

Equation \eqref{eq:f_time} predicts that cooler cores (lower $T_{\rm eq}$) accrete more mass. Using Equation \eqref{eq:mass_atm} we attribute this larger mass to the large Bondi radius $M_{\rm atm}\sim\rho_{\rm rcb} R_{\rm B}^3\propto\rho(R_c)T_{\rm eq}^{-1/2}$. The central density, on the other hand, increases with temperature $\rho(R_c)\propto T_{\rm eq}^{1/4}$, as seen by combining Equations \eqref{eq:enr_atm} and \eqref{eq:lum}. This distinction (which is sensitive to the opacity power-law, as discussed below) between dense and puffed-up atmospheres will be useful in interpreting the results of post-dispersal mass-loss, which we discuss in Section \ref{sec:dispersal}.

\section{Evolution Following Disk Dispersal}\label{sec:dispersal}

In Section \ref{sec:accretion} we affirmed that super Earth atmospheres grow to a few \% in mass until the nebula disperses after a few Myr \citep{Mamajek2009,WilliamsCieza2011,Alexander2014}. Here we discuss the evolution of these atmospheres after (and during) disk dispersal.

The post dispersal evolution is characterized by two processes: cooling and mass loss. In this phase cooling is equivalent to shrinking, and it occurs on a timescale $t/t_{\rm disk}= (R_{\rm rcb}/R_{\rm B})^{-2(3\gamma-4)/(\gamma-1)}$ according to Equation \eqref{eq:time_mass}, if $M_{\rm atm}$ is constant. We emphasize that the shrinking radius of the RCB $R_{\rm rcb}$ is a good approximation for the radius of the planet, since the scale height of the radiative layer is much smaller $h/R_{\rm rcb}\sim R_{\rm rcb}/R_{\rm B}<1$.
The above cooling timescale estimate is valid until $R_{\rm rcb}\sim R_c$, and our approximation of $r\ll R_{\rm rcb}$ in the inner envelope (see Section \ref{sec:accretion}) breaks down.
However, as we show in Section \ref{subsec:mass_loss}, thick envelopes shed part of their mass, and shrink to the thin regime on shorter timescales, so their cooling can be ignored. Before we discuss mass loss though, we analyze the structure of thin atmospheres.

\subsection{Thin Atmosphere}\label{subsec:thin_regime}

We now define $R_{\rm rcb}$ more generally as $R_{\rm rcb}\equiv R-R_c$, where $R$ is the radius of the planet (or the RCB). This definition coincides with the previous one for the thick atmosphere regime $R_{\rm rcb}\gg R_c$, and allows us to study thin envelopes with $R\sim R_c$, or equivalently $R_{\rm rcb}\lesssim R_c$, which are observationally interesting \citep{Lopez2012}. For a thin atmosphere Equation \eqref{eq:t_convective} can be approximated as
\begin{equation}\label{eq:tmp_thin}
\frac{T(r')}{T_{\rm rcb}}\approx 1+\frac{R_{\rm B}'}{R_c^2}r'\approx\frac{R_{\rm B}'}{R_c^2}r',
\end{equation}
with $r'$ measuring the distance from the RCB ($0<r'<R_{\rm rcb}$), and where we can ignore the thin isothermal layer of width $r'\sim R_c^2/R_{\rm B}'$ if the atmosphere is not ultra thin $R_{\rm rcb}/R_c>R_c/R_{\rm B}'$ (ultra thin atmospheres of exoplanets are currently observationally irrelevant).
Equation \eqref{eq:tmp_thin} immediately demonstrates a fundamental difference between the thin and thick regimes. A rocky core surrounded by a thick atmosphere does not cool, and stays at a constant temperature $T/T_{\rm rcb}=R_{\rm B}'/R_c$ (see Section \ref{subsec:accretion_cooling}). A thin envelope, on the other hand, allows the rocky core beneath it to cool, and the core's temperature decreases as the atmosphere shrinks $T/T_{\rm rcb}=R_{\rm B}'R_{\rm rcb}/R_c^2$. Therefore, in contrast to the thick regime, we expect the rocky core in the thin case to play a role in the energy budget of the cooling envelope, as we show below.

The density in the convective region is given by $\rho(r')/\rho_{\rm rcb}=(T(r')/T_{\rm rcb})^{1/(\gamma-1)}$. By integrating the density profile and using Equation \eqref{eq:tmp_thin}, we obtain the atmospheric mass
\begin{equation}\label{eq:matm_thin}
M_{\rm atm}=\frac{\gamma-1}{\gamma}4\pi R_c^2\rho_{\rm rcb}R_{\rm rcb}\left(\frac{R_{\rm B}'R_{\rm rcb}}{R_c^2}\right)^{1/(\gamma-1)}.
\end{equation}
Similarly to Equation \eqref{eq:de_dm}, and using the temperature and density profiles given by Equation \eqref{eq:tmp_thin}, we calculate the total (gravitational and thermal) energy available for cooling
\begin{equation}\label{eq:enr_thin}
E=gR_{\rm rcb}\left(\frac{\gamma}{2\gamma-1}M_{\rm atm}+\frac{1}{\gamma}\frac{\gamma-1}{\gamma_c-1}\frac{\mu}{\mu_c}M_c\right),
\end{equation} 
with $g\equiv GM_c/R_c^2$ denoting the surface gravity. The first term in Equation \eqref{eq:enr_thin} represents the energy of the gaseous envelope, while the second term accounts for the cooling of the rocky core, as discussed above. $\mu_c$ and $\gamma_c$ mark the rocky core's molecular weight and adiabatic index, respectively. For simplicity, we take $\gamma_c=4/3$, in accordance with the Dulong-Petit law. We see from Equation \eqref{eq:enr_thin} that although the core is more massive than the envelope, it may contain fewer particles due to its larger molecular weight, and therefore may have a lower energy capacity. We note that Equation \eqref{eq:enr_thin} assumes that the core remains molten, and therefore roughly isothermal. However, if the atmosphere is thin enough, the temperature at the atmosphere-core boundary decreases, and an insulating solid crust forms. Nonetheless, in this work we focus on sufficiently massive planets and atmospheres, for which the formation of a solid crust can be ignored up to late times.

By combining Equations \eqref{eq:lum}, \eqref{eq:matm_thin}, and \eqref{eq:enr_thin} we write an evolution equation ($L=-\dot{E}$) for the thin regime and obtain the time it takes the atmosphere to shrink to a width $R_{\rm rcb}$, assuming $M_{\rm atm}$ remains constant
\begin{equation}\label{eq:time_rad_thin}
\begin{split}
&t=\frac{1}{256\pi^2}\frac{3\gamma^2}{\gamma-1}\frac{\kappa M_{\rm atm}}{\sigma T_{\rm rcb}^3 R_c^4}\frac{k_{\rm B}}{\mu}\left(\frac{R_{\rm B}'R_{\rm rcb}}{R_c^2}\right)^{-1/(\gamma-1)}\\ &\times\left(\frac{\gamma}{2\gamma-1}M_{\rm atm}+\frac{1}{\gamma}\frac{\gamma-1}{\gamma_c-1}\frac{\mu}{\mu_c}M_c\right).
\end{split}
\end{equation}
Equation \eqref{eq:time_rad_thin} coincides with the thick regime Equation \eqref{eq:time_mass} for $R_{\rm rcb}\sim R_c$, up to numerical coefficients and the heat capacity of the core. According to Equation \eqref{eq:time_rad_thin} envelopes shrink with time in the thin regime as $t\propto R_{\rm rcb}^{-1/(\gamma-1)}$.

\subsection{Spontaneous Mass Loss}\label{subsec:mass_loss}

In previous sections we calculated the radius evolution of the planet after disk dispersal, assuming that the envelope mass remains constant. In particular, we found that the cooling time to envelope thickness $R_{\rm rcb}$ is $t\propto R_{\rm rcb}^{-1}$ for $R_{\rm rcb}>R_c$ ({\it thick regime}) and $t\propto R_{\rm rcb}^{-5/2}$ for $R_{\rm rcb}<R_c$ ({\it thin regime}), assuming $\gamma=7/5$.
However, as we show below, mass loss is inherent to post-dispersal evolution. By comparing the mass-loss timescale and the cooling time, we find below that thick atmospheres are always governed by mass loss, while thin envelopes can be either mass-loss or cooling dominated, depending on their mass.

When the disk disperses, the gas density $\rho_d\to 0$. If the dispersal process is faster than the envelope cools, then according to Equation \eqref{eq:radius_rho_d}, we expect $R_{\rm rcb}$ to decrease, while $\rho_{\rm rcb}$ remains constant, since the energy does not change, and the energy of thick atmospheres is determined only by $\rho_{\rm rcb}$, not $R_{\rm rcb}$, as seen in Equation \eqref{eq:enr_atm}. Consequently, we expect the atmosphere to decrease in mass, according to Equation \eqref{eq:mass_atm}, at least until the thin regime is reached. However, even if the loss of pressure support from the ambient disk (causing the mass loss) is immediate (see Section \ref{sec:transition}), the atmosphere adjusts itself to the changing boundary condition on a finite timescale. 

A commonly used criterion to determine the mass loss rate is the energy limited argument. The basic picture is that while the gas at the Bondi radius can escape to vacuum (the molecule escape rate at the Bondi radius is discussed below), in order for mass loss to continue, gas from deeper in the planet's potential well has to reach the Bondi radius and replenish the escaping gas. However, as gas expands adiabatically to reach the Bondi radius, its temperature drops to zero after expanding only a scale height $h=R_{\rm rcb}^2/R_{\rm B}\ll R_{\rm rcb}$ (simple adiabatic atmosphere solution). Therefore, in order to elevate the atmosphere out of the planet's potential well, a constant supply of energy is required. While most studies focus on ionizing stellar photons, which we consider in Section \ref{sec:photo_evap}, as the energy source \citep{MurrayClay2009,Lopez2012, OwenJackson2012, OwenWu2013}, we first examine the envelope's own cooling luminosity \citep[see also][]{OwenWu2015} as an energy source driving mass loss \citep[see also][who consider the core's luminosity, which we also discuss below]{IkomaHori2012}.

As we explain in Section \ref{sec:accretion}, for $\gamma=7/5$ the mass of the atmosphere is concentrated at the outside, at $r\sim R_{\rm rcb}$ (see discussion below for different values of $\gamma$). Therefore, the amount of energy needed to blow most of the atmosphere away is $E_{\rm evap}\sim GM_cM_{\rm atm}/R_{\rm rcb}$, where we approximate $R_{\rm rcb}\ll\min (R_{\rm B},R_{\rm H})$ for the shrinking phase after disk dispersal. The energy that the envelope loses by cooling, on the other hand, is concentrated in the inside, and is given by Equation \eqref{eq:energy_mass_radius}. By comparing the two results, we find that the cooling inner envelope can blow off the outer atmosphere (which contains most of the mass) without changing its energy by much. Consequently, the planet loses mass while its energy, and therefore $\rho(R_c)$ and $\rho_{\rm rcb}$, remain constant, due to Equation \eqref{eq:enr_atm}. The planet's radius, according to Equation \eqref{eq:mass_atm}, shrinks as it loses mass. Explicitly, the ratio of evaporation to cooling timescales is given by
\begin{equation}\label{eq:ratio_evap_cool_thick}
\frac{t_{\rm evap}}{t_{\rm disk}}=\frac{t_{\rm evap}}{t_{\rm cool}}=\frac{E_{\rm evap}}{E_{\rm cool}}\sim\left(\frac{R_{\rm rcb}}{R_c}\right)^{-(3-2\gamma)/(\gamma-1)},
\end{equation}
where the first equality ($t_{\rm cool}=t_{\rm disk}$) is trivial at disk dispersal (when $R_{\rm rcb}=R_{\rm rcb}^0\lesssim R_{\rm B}$), since the planet has cooled for $t_{\rm disk}$. This equality continues to be true at later times (when $R_{\rm rcb}<R_{\rm rcb}^0$) since both the energy, and the luminosity, given by Equation \eqref{eq:lum}, depend only on $\rho_{\rm rcb}$ (and not $R_{\rm rcb}$), which remains constant during the mass loss phase. Consequently, $t_{\rm cool}$ is constant during the mass loss. Equation \eqref{eq:ratio_evap_cool_thick} shows that envelopes shed mass until they enter the thin regime $R_{\rm rcb}\sim R_c$, at a time comparable with $t_{\rm disk}$. Since, according to Equation \eqref{eq:mass_atm}, $M_{\rm atm}\propto R_{\rm rcb}^{1/2}$ for $\gamma=7/5$, and $R_c/R_{\rm B}\sim 0.1$, atmospheres retain roughly 30\% of their initial mass when they enter the thin regime. 

Explicitly, by combining the accreted atmosphere mass at disk dispersal, given by Equation \eqref{eq:f_time}, with the factor of $\sim(R_c/R_{\rm B})^{1/2}$ due to the outer envelope shedding, we derive the mass fraction of the atmosphere when it reaches the thick-thin transition:
\begin{equation}\label{eq:f_semi_thin}
	f_{\rm semi-thin}\approx 0.01\left(\frac{M_c}{M_\Earth}\right)^{0.44}\left(\frac{T_{\rm eq}}{10^3\textrm{ K}}\right)^{0.25}\left(\frac{t_{\rm disk}}{1\textrm{ Myr}}\right)^{0.5}.
\end{equation} 
Equation \eqref{eq:f_semi_thin} demonstrates that although it is easier to acquire a heavy atmosphere at larger orbital separations \citep[see, e.g., Section \ref{sec:accretion}, or][]{LeeChiang2015}, hotter cores (in close orbits) retain a heavier atmosphere after the disk dispersal and the shedding of the bloated (and weakly bound) outer envelope, since they reach higher densities, as explained in Section \ref{sec:accretion}, and $M_{\rm atm}\sim\rho(R_c) R_c^3\propto T_{\rm eq}^{1/4}$ at this stage (a strong increase of the opacity with temperature can change this conclusion, as discussed below). 
 
For $\gamma<4/3$, as effectively chosen by \citet{LeeChiang2015}, the mass and the energy are both concentrated in the inside, so $E_{\rm evap}\sim E_{\rm cool}\sim GM_cM_{\rm atm}/R_c$, and the atmosphere mass is simply $M_{\rm atm}\sim \rho(R_c)R_c^3$. In this case, it is obvious that the loosely bound outer envelope is shed before the inner layers cool significantly, on a timescale $\lesssim t_{\rm disk}$. Therefore, regardless of $\gamma$, atmospheres shed their outer layers and shrink to $R_{\rm rcb}
\sim R_c$ after the nebula vanishes. However, if $\gamma<4/3$, these outer layers do not contain significant mass, and atmospheres retain most of their initial mass when they reach the thin regime. Quantitatively, atmospheres retain $1-2^{-(4-3\gamma)/(\gamma-1)}\approx 75\%$ of their mass in this case \citep[for $\gamma=1.2$ chosen by][]{LeeChiang2015}. 
The subsequent thin-regime evolution does not depend qualitatively on $\gamma$, as discussed below. 

Once an envelope sheds its outer layers and enters the thin regime, the energy required to blow away its atmosphere is given by $E_{\rm evap}\sim GM_cM_{\rm atm}/R_c=M_{\rm atm}gR_c$. In order to check if the atmosphere continues to shed mass in the thin regime, we distinguish between two cases, according to Equation \eqref{eq:enr_thin}. {\it Heavy atmospheres}, with $M_{\rm atm}/M_c\gtrsim\mu/\mu_c$, regulate their own cooling, while the heat capacity of {\it light atmospheres}, with $M_{\rm atm}/M_c\lesssim\mu/\mu_c$, is negligible, and they are controlled by the cooling rocky core underneath (which can cool only once the atmosphere is thin, see Section \ref{subsec:thin_regime}). Here $M_{\rm atm}$ is the remaining mass of the semi-thin atmosphere, not the initial mass. Assuming an Earth-like composition for the rocky core, the critical gas to solid ratio which distinguishes between heavy and light atmospheres is $f\approx 5\%$.

\subsubsection{Heavy Atmospheres}\label{sec:heavy}

For heavy atmospheres, initially (when $R_{\rm rcb}=R_c$) the cooling rate and mass loss rate are equal, since $E_{\rm cool}\sim E_{\rm evap}=M_{\rm atm}gR_c$. However, as $R_{\rm rcb}$ decreases (due to cooling) the cooling timescale becomes shorter by a factor $R_{\rm rcb}/R_c<1$ in comparison with the mass-loss time, according to Equation \eqref{eq:enr_thin}, so we can consider cooling at a constant envelope mass, as in Section \ref{subsec:thin_regime}.

The envelope contracts with time as $t\propto R_{\rm rcb}^{-5/2}$. However, gas envelopes cannot compress indefinitely, and they reach a maximum density of $\rho_{\rm max}\sim\mu/a_0^3$, where $a_0$ is the Bohr radius. When heavy envelopes reach the thin regime, their density is
\begin{equation}
\rho\sim\frac{3}{4\pi}\frac{M_{\rm atm}}{7R_c^3}\gtrsim\frac{3}{4\pi}\frac{\mu}{\mu_c}\frac{M_c}{7R_c^3}\sim\frac{\mu}{\mu_c}\frac{\rho_c}{7}\sim\frac{\rho_{\rm max}}{7},
\end{equation}
with $\rho_c$ marking the rocky core's density (not to be confused with $\rho(R_c)$, which is the gas density at $r=R_c$), and assuming that $\rho_c\sim\mu_c/a_0^3$ (a crude approximation of roughly the same radius for all atoms, due to electron screening of the nuclear charge). We conclude that heavy envelopes shrink by a maximal factor $\approx 7$ before reaching their maximum gas density. This contraction lasts for $7^{5/2}\cdot t_{\rm disk}\approx 10^2\cdot t_{\rm disk}\sim 1\textrm{ Gyr}$ at most, since $t_{\rm disk}$ is the cooling time for $R_{\rm rcb}\sim R_c$. Thus, atmospheres of Gyr old planets are no longer contracting, and their density is $\approx\rho_{\rm max}$ \citep[we ignore here possible inflation mechanisms similar to those invoked for inflated hot Jupiters; see, e.g.,][]{ValenciaPu2015}.

\subsubsection{Light Atmospheres}\label{subsec:light}

Light atmospheres, on the other hand, start the thin regime ($R_{\rm rcb}=R_c$) with mass loss timescales which are shorter than their cooling time, according to Equation \eqref{eq:enr_thin}. As these envelopes lose mass, $R_{\rm rcb}$ remains constant (since the energy does not decrease), while $\rho_{\rm rcb}$ decreases, according to Equation \eqref{eq:matm_thin}. As a result, the energy required for evaporation only decreases (since it is $\propto M_{\rm atm}$), while the cooling energy, which is dominated by the rocky core, remains constant. In this way, light atmospheres are lost completely.

An additional timescale limiting the atmospheric loss is due to the finite escape rate of molecules from the Bondi radius \citep[see also][]{OwenWu2015} $\dot{M}=4\pi R_{\rm B}^2\rho(R_{\rm B})c_s$, which limits the atmosphere loss time to $t\sim M_{\rm atm}/\dot{M}$, explicitly
\begin{equation}\label{eq:escape_sound}
t\sim\frac{R_{\rm B}'}{c_s}\left(\frac{R_{\rm rcb}}{R_{\rm B}'}\right)^{(3\gamma-4)/(\gamma-1)}\exp{\left(\frac{R_{\rm B}}{R_{\rm rcb}}-1\right)},
\end{equation}
where we use Equation \eqref{eq:mass_atm} and find $\rho(R_{\rm B})$ using Equation \eqref{eq:radius_rho_d}. This timescale is longer than the planet's age, $t\sim\textrm{Gyr}$, when the planet reaches the thin regime, if 
\begin{equation}\label{eq:escape_sound_age}
\frac{R_{\rm B}}{2R_c}\gtrsim1+\ln\left[\frac{tc_s}{R_{\rm B}'}\left(\frac{R_{\rm B}'}{R_c}\right)^{(3\gamma-4)/(\gamma-1)}\right]\approx 30.
\end{equation}
We rewrite Equation \eqref{eq:escape_sound_age} as a condition on the mass and equilibrium temperature of the planet:
\begin{equation}\label{eq:cond_bondi}
\frac{M_c}{M_\Earth}\gtrsim 6.3\left(\frac{T_{\rm eq}}{10^3\textrm{ K}}\right)^{4/3}.
\end{equation}

In summary, gaseous envelopes shed their outer layers and reach a radius $\sim R_c$ on a timescale of the disk lifetime. At this stage, the cooling of the rocky core evaporates envelopes lighter than $M_{\rm atm}/M_c\approx 5\%$, though this evaporation may take longer than the current age of the planet if it is massive or cold enough (see Figure \ref{fig:obs}). We note that photo-evaporation due to high energy stellar photons does not obey the above escape rate limit at the Bondi radius, as explained in Section \ref{sec:photo_evap}. 

\subsection{Transitional Disks}\label{sec:transition}

We have assumed above that the depletion of the disk is faster than the time it takes the atmosphere to adjust to the vanishing boundary condition $\rho_d\to 0$. Quantitatively, we assumed that $t_{\rm trans}<t_{\rm evap}$, with $t_{\rm trans}$ denoting the transition timescale for the disk to disperse. This assumption is justified by noticing that, considering Equation \eqref{eq:ratio_evap_cool_thick}, $1/3\lesssim t_{\rm evap}/t_{\rm disk}\lesssim 1$, while, by definition, $t_{\rm trans}<t_{\rm disk}$, and observations suggest that $t_{\rm trans}/t_{\rm disk}\approx 0.1$ \citep{Alexander2014}. We therefore conclude that the finite disk dispersal time is irrelevant for our above analysis, since the mass-loss bottleneck is the energy release anyhow.

Nonetheless, it is interesting to consider the gradual decrease of the nebula's density $\rho_d$ over time. Because the gas accretion depends on $\rho_d$ logarithmically (see Section \ref{sec:accretion}), cores can accumulate gas envelopes even in a depleted nebula, as suggested by \citet{LeeChiang3}. Accretion from a gas-poor disk might be important if collisions of isolation masses (that occur only once the gas density is low enough so it cannot damp eccentricity excitations) remove their initial atmospheres \citep{InamdarHilke}. The atmosphere growth in this case is determined by a competition between the cooling of the envelope and the dispersal of the disk.

Quantitatively, we parametrize the disk dispersal over time as $\rho_d(t)=\rho_d^0\exp[-(t/t_{\rm trans})^\alpha]$, allowing for an exponential depletion as modeled by \citet{IkomaHori2012}, as well as more or less gradual processes \cite[see, e.g.][who incorporate a linear decrease with time]{Rogers2011}. By combining Equations \eqref{eq:enr_atm} and \eqref{eq:lum} we find that the atmosphere's density increases as $\rho_{\rm rcb}\propto t^{1/2}$, while its radius decreases approximately as $R_{\rm rcb}\propto t^{-\alpha}$, using Equation \eqref{eq:radius_rho_d}, and assuming that the nebula has already depleted significantly. Consequently, using Equation \eqref{eq:mass_atm}, the mass of the atmosphere grows as $M_{\rm atm}\propto R_{\rm rcb}^{1/2}\rho_{\rm rcb}\propto t^{(1-\alpha)/2}$. We conclude that if the dispersal is gradual enough ($\alpha<1$) then an atmosphere can grow while the disk is being depleted. This analysis is valid only as long as $R_{\rm rcb}>R_c$. In the thin regime, Equation \eqref{eq:radius_rho_d} implies that $\rho_{\rm rcb}(t)\propto\rho_d(t)$, preventing further atmosphere growth.

\section{Mass Loss by Photo-evaporation}\label{sec:photo_evap}

In Section \ref{subsec:mass_loss} we examined mass loss which is powered by the heat from the inner envelope, or the rocky core. In this section we address another energy source --- ionizing stellar photons. Mass lass by photo-ionization has been studied extensively, both in the context of hot Jupiters and super Earths \citep{Baraffe2004,Baraffe2005,Baraffe2006,Hubbard2007a,Hubbard2007b,MurrayClay2009,Jackson2010,Valencia2010,Lopez2012, OwenJackson2012, OwenWu2013}. The basic picture is that ionizing photons release energetic electrons which in turn heat the gas to high temperatures, above the escape velocity. If the cooling of the gas is slow enough, the high temperature gas escapes the planet's potential well. We emphasize that the stellar continuum radiation cannot heat the gas above the equilibrium temperature, which is lower than the escape velocity (below the Bondi radius). Therefore, the continuum radiation can only provide energy to elevate the gas to the Bondi radius, in a similar manner to the cooling luminosity, as described in Section \ref{subsec:mass_loss} \citep[see also][]{OwenWu2015}. In fact, the stellar continuum radiation which heats the gas has the same magnitude as the cooling luminosity $\sim\sigma T_{\rm rcb}^4R_{\rm rcb}^2/\tau$.

Here, we adopt the popular simplified energy-limited model for the photo-evaporation \citep[see, e.g.][and references within]{Lopez2012,OwenWu2013}, which linearly connects the high-energy flux to the gravitational energy of the escaping mass. Explicitly, we assume that the photo-evaporating flux can be written as $L=4\pi R_{\rm rcb}^2\sigma T_{\rm rcb}^4\epsilon$, with $\epsilon\sim 10^{-4}$ taking into account both the evaporation efficiency $\sim 0.1$ \citep[see, e.g.,][]{OwenWu2013} and the small fraction of ionizing radiation out of the total bolometric stellar flux, which is taken to be a constant $\sim 10^{-3}$ for the first $t_{\rm UV}\sim 100\textrm{ Myr}$, and then decreases with time as $t^{-1.25}$ \citep[see, e.g.][and references within]{Jackson2012,Lopez2012,OwenJackson2012}. The UV evaporation timescale can therefore be written as
\begin{equation}\label{eq:photo_evap}
t_{\rm evap}=\frac{M_{\rm atm}gR_c}{4\pi R_c^2\sigma T_{\rm rcb}^4\epsilon},
\end{equation}
where we focus on the {\it thin atmosphere} regime, since thick atmospheres shed mass even without UV radiation. 

We emphasize that even if the condition $t_{\rm evap}>t_{\rm cool}$ is satisfied for $R_{\rm rcb}=R_c$, since the cooling time increases as $t_{\rm cool}\propto R_{\rm rcb}^{-1/(\gamma-1)}$ for a constant mass, while $t_{\rm evap}$ remains constant (does not depend on $R_{\rm rcb}$ or time for $t<t_{\rm UV}$), planets will lose their mass at some point. The only way to retain the atmosphere is to ensure that $t_{\rm evap}>t_{\rm UV}$. After $t_{\rm UV}$, the ratio of evaporation timescale to age scales as $t_{\rm evap}/t\propto t^{0.25}$, so if an atmosphere survived to $t_{\rm UV}$, it will retain most of its mass afterwards.

Although Equation \eqref{eq:photo_evap} shows that massive envelopes may survive photo-evaporation, since $t_{\rm evap}\propto M_{\rm atm}$, it is important to check whether such massive envelopes could have formed during the disk's lifetime $t_{\rm disk}$. A similar concern was raised by \citet{Lopez2012}, who find that some of the Kepler-11 planets had to start with tens of percents of their mass in gas in order for some of the atmosphere to survive photo-evaporation. Such massive atmospheres are problematic for two reasons --- planets might have lacked the time to accrete them (see Section \ref{sec:accretion}), and the self gravity of these envelopes brings them close to the runaway accretion regime. We address the problem of nebular gas accretion followed by photo-evaporation self-consistently, by writing the evaporation timescale in terms of the disk's lifetime  
\begin{equation}\label{eq:epsilon_tau}
\frac{1}{\epsilon\tau}\sim\frac{t_{\rm evap}}{t_{\rm cool}(R_c)}\gtrsim\frac{t_{\rm UV}}{t_{\rm disk}}\approx 10,
\end{equation}
where $\tau\sim\kappa\rho_{\rm rcb} h\sim\kappa \rho_{\rm rcb} R_c^2/R_{\rm B}$ is the optical depth at the RCB. Equation \eqref{eq:epsilon_tau} simply states that since the evaporation and cooling energies are the same at $R_{\rm rcb}=R_c$ (assuming the {\it heavy regime}), the ratio between the timescales is given by the ratio between the evaporation flux $\sigma T_{\rm rcb}^4\epsilon$ and the cooling flux $\sigma T_{\rm rcb}^4/\tau$. $t_{\rm cool}(R_c)$ is also equal to $t_{\rm disk}$ (see Section \ref{subsec:mass_loss}). By substituting the density from Equation \eqref{eq:mass_atm} or \eqref{eq:matm_thin}, we obtain the condition
\begin{equation}\label{eq:photo_evap_condition}
\frac{\epsilon\kappa M_{\rm atm}}{R_cR_{\rm B}}\left(\frac{R_c}{R_{\rm B}'}\right)^{1/(\gamma-1)}\lesssim \frac{t_{\rm disk}}{t_{\rm UV}}\approx 0.1.
\end{equation}
It is initially counterintuitive that a small envelope mass is required to survive evaporation, since $t_{\rm evap}\propto M_{\rm atm}$. The reason is that $t_{\rm disk}\propto M_{\rm atm}^2$, as explained in Section \ref{sec:accretion}, so for given disk and UV activity times, lighter envelopes survive evaporation because $t_{\rm evap}/t_{\rm disk}\propto M_{\rm atm}^{-1}$. We substitute the atmosphere mass from Equation \eqref{eq:f_semi_thin}, which takes into account self-consistently both the accretion and the spontaneous shedding of the outer gas envelope, and rewrite Equation \eqref{eq:photo_evap_condition} as
\begin{equation}\label{eq:cond_uv}
	\frac{M_c}{M_\Earth}\gtrsim 7.7\left(\frac{T_{\rm eq}}{10^3\textrm{ K}}\right)^{2.22},
\end{equation}
showing that cores have to be massive or ``cold'' (low equilibrium temperature) to keep an atmosphere. However, as discussed in Section \ref{sec:accretion} \citep[see also][]{LeeChiang2015}, cores which are too massive or cold are at the risk of runaway gas accretion. The overlap between the two conditions is discussed in Section \ref{sec:obs}. 

It is worthwhile to repeat the derivation of Equation \eqref{eq:cond_uv}, which is one of our main results, for the case of $\gamma<4/3$, due to Hydrogen dissociation, as suggested by \citet{LeeChiang2015} and \citet{Piso2015}. The calculation in this case takes a simpler form since the mass of the atmosphere is concentrated in the inside, as the energy, so $E\sim GM_cM_{\rm atm}/R_c$, and there is no significant spontaneous mass loss following the disk's dispersal (see Section \ref{subsec:mass_loss}). By repeating the calculation of $M_{\rm atm}$ with $\gamma=1.2$ \citep{LeeChiang2015}, and substituting in Equation \eqref{eq:photo_evap_condition}, we find that the qualitative shape of the critical curve for mass retention, described by Equation \eqref{eq:cond_uv}, does not change significantly. Specifically, we find that $M_c\propto T_{\rm eq}^{1.9}$ in this case.

Equation \eqref{eq:epsilon_tau} demonstrates that the RCB density $\rho_{\rm rcb}$ has to be low, for the optical depth $\tau\sim\kappa\rho_{\rm rcb} h$ to be low, enabling accretion to be efficient in comparison with evaporation. However, as explained in Section \ref{sec:accretion}, $\rho_{\rm rcb}$ has a lower limit dictated by the surrounding nebula $\rho_{\rm rcb}>\rho_d$ (note that $\rho_{\rm rcb}$ does not change during the thick atmosphere mass loss phase, since the energy, which is determined solely by $\rho_{\rm rcb}$ remains constant, as explained in Section \ref{subsec:mass_loss}). Close enough to the star, the gas densities are high
\begin{equation}
\rho_d=10^{-6}\textrm{ g cm}^{-3}\left(\frac{T_{\rm eq}}{10^3\textrm{ K}}\right)^{5.5},
\end{equation}
assuming the MMSN model \cite[see Section \ref{subsec:adiabatic} and][]{Hayashi1981}. By substituting this minimal density in Equation \eqref{eq:epsilon_tau} we obtain an additional condition relating the mass and temperature of planets with atmospheres
\begin{equation}\label{eq:min_rho_symbol}
\epsilon\kappa\rho_d\frac{R_c^2}{R_{\rm B}}\lesssim\frac{t_{\rm disk}}{t_{\rm UV}},
\end{equation}
or quantitatively
\begin{equation}\label{eq:min_rho}
\frac{M_c}{M_\Earth}\gtrsim 10^{-4}\left(\frac{T_{\rm eq}}{10^3\textrm{ K}}\right)^{13}.
\end{equation} 
Equation \eqref{eq:min_rho} also requires low temperatures (for low disk densities) and high masses (for a small scale height $h=k_{\rm B}T/\mu g$, due to the strong gravity). However, by comparing Equations \eqref{eq:cond_uv} and \eqref{eq:min_rho}, we find that the disk density constraint, given by Equation \eqref{eq:min_rho}, is relevant only for $T_{\rm eq}\gtrsim 3000\textrm{ K}$, rendering it irrelevant for the set of observations (see Section \ref{sec:obs}). This result remains true even for denser nebulae (up to a factor of $10^2$), as seen in Equation \eqref{eq:min_rho_symbol}. 

\subsection{Migration}\label{subsec:migration}

We have so far assumed that gas accretion and UV evaporation occur at the same distance from the star, which allowed us to relate the evaporation time to the nebular accretion time, and to constrain the formation location and planet mass. Can separating the accretion from evaporation, due to migration during the nebula's presence, relax the constraints on the formation scenario? 
According to Equation \eqref{eq:f_semi_thin}, hotter cores (on closer orbits) retain a more massive atmosphere after the nebula disperses.
Therefore, we intuitively expect, due to Equation \eqref{eq:photo_evap}, that these heavy atmospheres will survive the UV evaporation at later times.

We now consider gas accretion at an initial distance from the star characterized by an equilibrium temperature $T_i$ for a time $\sim t_{\rm disk}$, followed by migration on a similar timescale (while gas is still present), and UV irradiation at a different orbit, characterized by $T_f$. In this case, using Equations \eqref{eq:lum} and \eqref{eq:photo_evap}, Equation \eqref{eq:epsilon_tau} changes to
\begin{equation}\label{eq:condition_migration_sym}
\left(\frac{T_i}{T_f}\right)^4\frac{1}{\epsilon\tau_i}\gtrsim\frac{t_{\rm UV}}{t_{\rm disk}},
\end{equation}
with $\tau_i$ denoting the optical depth (which is relevant for accretion) at $T_i$. Correspondingly, the condition for a rocky core to retain its atmosphere, given by Equation \eqref{eq:cond_uv}, changes to
\begin{equation}\label{eq:condition_migration}
\frac{M_c}{M_\Earth}\gtrsim 7.7 \left(\frac{T_i}{10^3\textrm{ K}}\right)^{2.22}\left(\frac{T_f}{T_i}\right)^{2.37}.
\end{equation}
Equation \eqref{eq:condition_migration} demonstrates that migration relaxes the critical mass constraint for atmosphere retention. Specifically, as expected, cores that migrated outward to a given $T_f\approx 10^3\textrm{ K}$ (the planet's current location) can have low critical masses $M_c\propto T_i^{-0.15}$ and still survive evaporation. 
The effect of migration takes a more general form $M_c\propto T_i^{-(0.5-b)/3.37}$, if we incorporate a temperature-dependent opacity $\kappa\propto T^b$ into Equation \eqref{eq:condition_migration_sym}. Consequently, if there is a strong increase of the opacity with temperature, inward migration during the gas disk's lifetime may relax the core mass constraint. The reason is that in this case colder cores accrete and retain more mass after the disk disperses, as seen by plugging the temperature-dependent opacity in Section \ref{sec:accretion}. Specifically, using Equations \eqref{eq:time_mass} and \eqref{eq:f_semi_thin} we find $f_{\rm semi-thin}\propto T_{\rm eq}^{(0.5-b)/2}$. 

\section{Comparison with Observations}\label{sec:obs}

In previous sections we explained how accretion and evaporation histories limit the population of planets which can evolve in-situ into low density super Earths. Specifically, in Section \ref{sec:accretion} and Equation \eqref{eq:f_time} we show that planets which are too massive or too cold explode into Jupiters \citep[see also][]{LeeChiang2015}, while in Sections \ref{subsec:light} and \ref{sec:photo_evap} we show that planets which are too light or too hot lose their atmosphere due to UV evaporation or cooling of the rocky core. In this section we compare the observed super-Earth population to this theoretically allowed ``Goldilocks'' regime.

\begin{figure}[tbh]
	\includegraphics[width=\columnwidth]{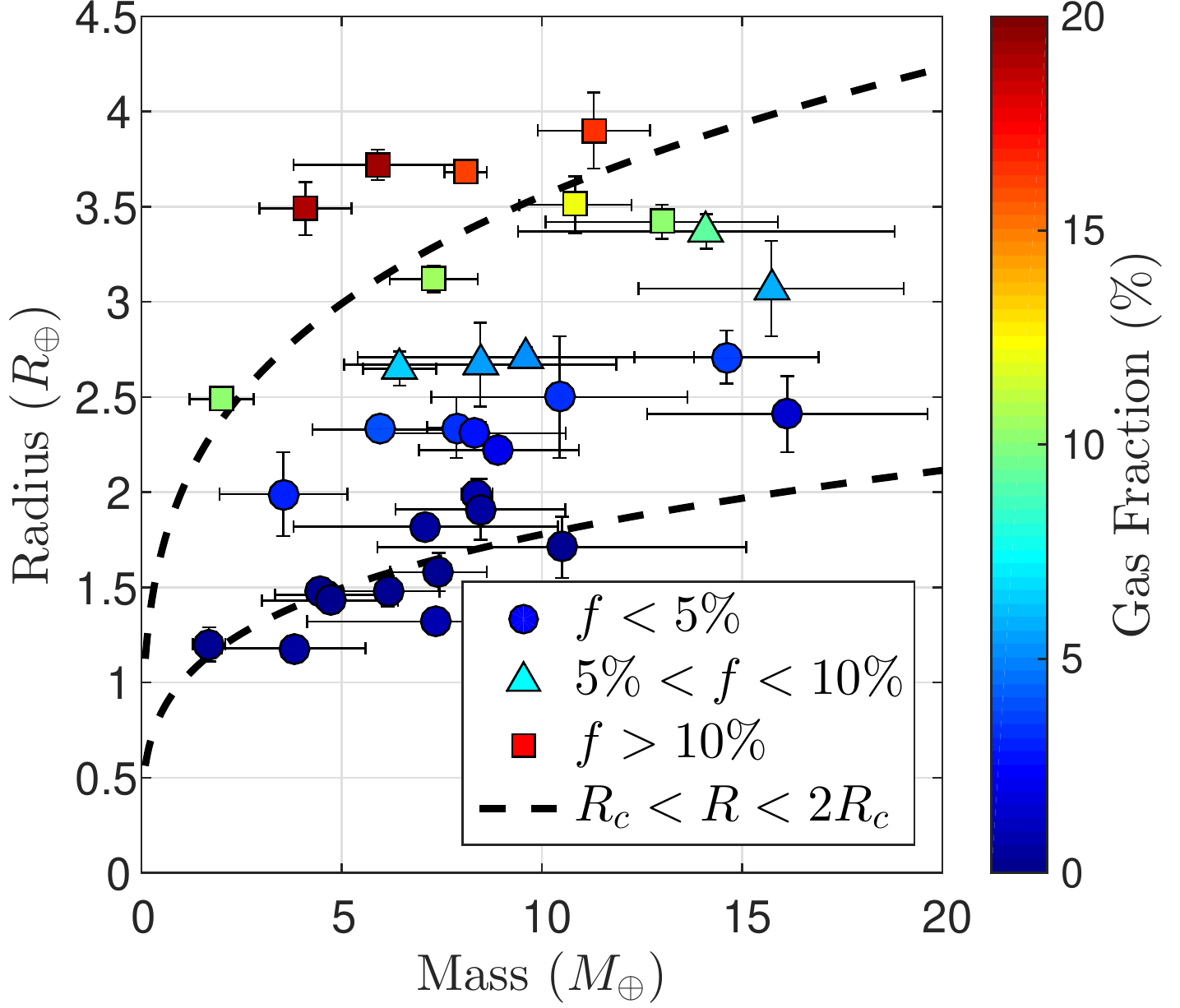}
	\caption{Observed super-Earth population (see text for details) from \citet{WeissMarcy2014}. The planets are grouped according to their gas mass fraction $f$, estimated by Equation \eqref{eq:obs_f}, with low-density planets marked by triangles ($5\%<f<10\%$) or squares ($f>10\%$). The planet markers are also color-coded according to $f$. The two dashed black lines mark the radius of the rocky core $R_c(M_c)$ and $2R_c(M_c)$. Planets with substantial atmospheres are expected to be found roughly between the two lines.  
		\label{fig:obs_rc}}
\end{figure}

\begin{figure}[tbh]
	\includegraphics[width=\columnwidth]{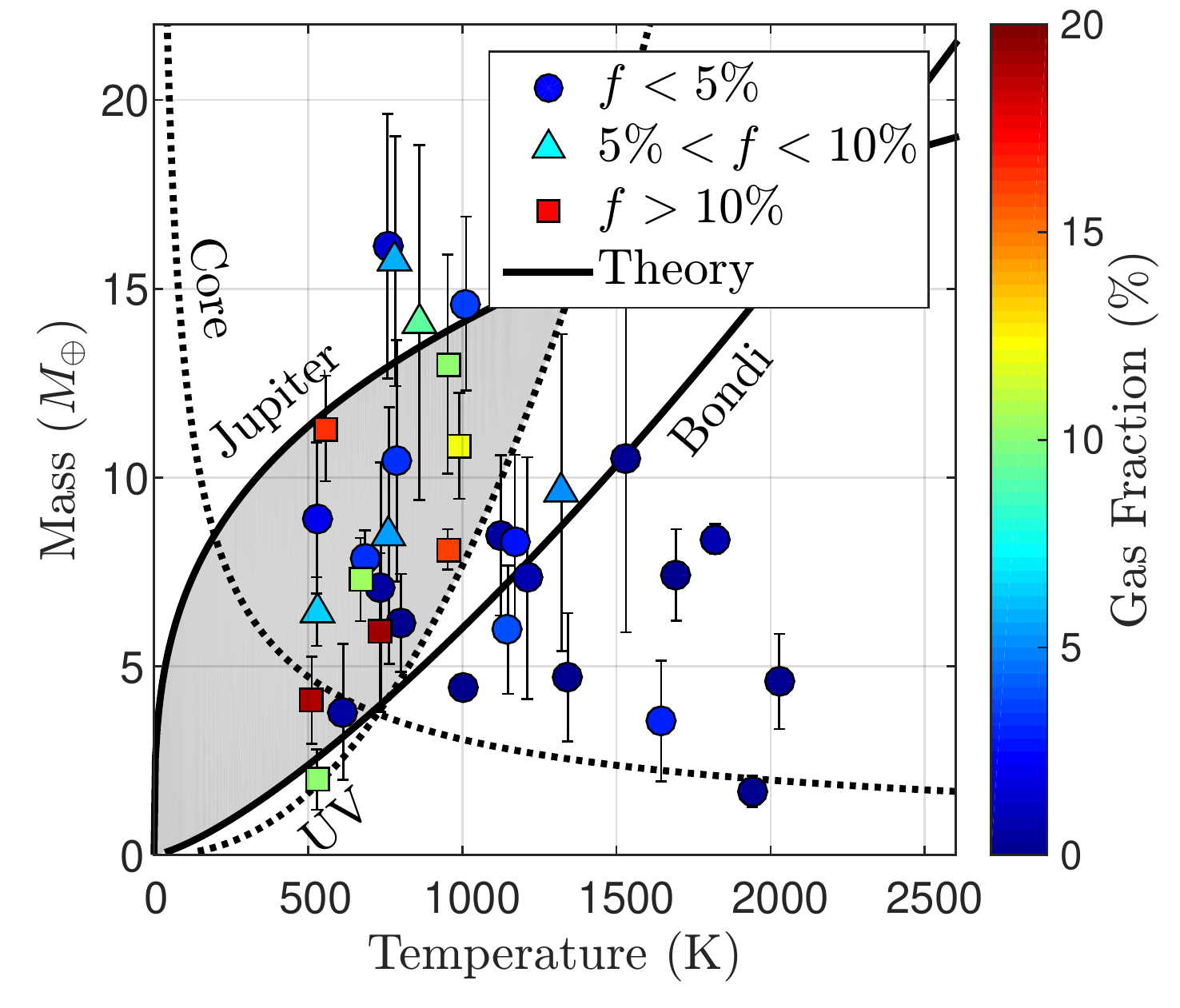}
	\caption{Observed super-Earth population (see text for details) from \citet{WeissMarcy2014}. The planets are grouped according to their gas mass fraction $f$, estimated by Equation \eqref{eq:obs_f}, with low-density planets marked by triangles ($5\%<f<10\%$) or squares ($f>10\%$). The planet markers are also color-coded according to $f$. The top solid line (``Jupiter'') is according to Equation \eqref{eq:f_time} with $t_{\rm disk}=10\textrm{ Myr}$ and $f=0.5$, while the dashed ``Core'' line is according to Equation \eqref{eq:f_semi_thin} with $f=5\%$. The bottom solid line (``Bondi'') follows Equation \eqref{eq:cond_bondi}, and the dashed ``UV'' line follows Equation \eqref{eq:cond_uv}. Inside the shaded area, planets manage to accrete and maintain gas envelopes without exploding into gas giants due to runaway accretion (see text for details). 
		\label{fig:obs}}
\end{figure}

In Figures \ref{fig:obs_rc} and \ref{fig:obs} we present scatter plots of the observed planets with $R<4R_\Earth$, orbital periods shorter than 100 days, and  an error of less than 50\% in mass from \citet{WeissMarcy2014}. We estimate the equilibrium temperature by $T/T_\Earth=(F/F_\Earth)^{1/4}$, with the flux (relative to Earth's) $F/F_\Earth$ estimated by \citet{WeissMarcy2014}, and with the Earth's equilibrium temperature $T_\Earth\approx 260\textrm{ K}$. The observed atmospheric mass fraction is estimated by
\begin{equation}\label{eq:obs_f}
f=\frac{\rho_{\rm max}}{\rho_c}\left[\left(\frac{R}{R_c}\right)^3-1\right],
\end{equation}
since Gyr old atmospheres are close to the maximum gas density (see Section \ref{sec:heavy}), which we evaluate using the equation of state of \cite{Nettelmann2008} as 
\begin{equation}\label{eq:rho_max}
\rho_{\rm max}\approx 0.5\textrm{ g cm}^{-3}\left(\frac{P}{\textrm{Mbar}}\right)^{0.4},
\end{equation}
with the typical atmosphere pressure given by
\begin{equation}\label{eq:pressure_atm}
P=\frac{M_{\rm atm}g}{4\pi R^2}\approx\textrm{Mbar}\left(\frac{M}{M_\Earth}\right)^2\left(\frac{R}{R_\Earth}\right)^{-4}f.
\end{equation}
We take into account the mild compression of the rocky core \citep[see, e.g.,][]{Valencia2006}, and estimate the rocky core's density and radius by $\rho_c/\rho_\Earth\approx(M_c/M_\Earth)^{1/4}$ and $R_c/R_\Earth\approx(M_c/M_\Earth)^{1/4}$, with $\rho_\Earth=5.5\textrm{ g cm}^{-3}$. Equation \eqref{eq:obs_f} with the approximation $M\approx M_c$ ($M$ is the observed mass) is a valid estimate for $f\ll 1$, and indeed all planets in our sample have estimated $f\leq20\%$. Our crude estimate of the gas fraction is in agreement (approximately) with more elaborate estimates, e.g., \citet{Lopez2012}. We emphasize that an approximate estimate of $f$ suffices, since we are mainly interested in distinguishing between purely rocky worlds and planets with substantial atmospheres (above a few \% in mass) which have a significant volume. We plot in Figure \ref{fig:obs_rc} the limits of the thin envelope regime $R_c<R\lesssim 2R_c$, which according to our model confines super Earths older than $\sim 10\textrm{ Myr}$. 

In Figure \ref{fig:obs} we present the allowed Goldilocks region in which super-Earths can hold on to substantial atmospheres, taking into account all the mass-loss processes mentioned in previous sections. Specifically, planets below the ``Core'' line in the figure are left with $f<5\%$ after disk-dispersal, and therefore lose their atmospheres in less than $\sim$ Gyr due to the cooling of the core, if they are also below the ``Bondi'' line (see Section \ref{subsec:light}). Planets below the ``UV'' line, on the other hand, lose their atmospheres due to UV irradiation, regardless of the ``Bondi'' line (see Section \ref{sec:photo_evap}). Planets above the ``Jupiter'' line explode into gas giants during the disk's lifetime (see Section \ref{sec:accretion}).

As demonstrated in Figure \ref{fig:obs}, our model defines a relatively narrow mass/temperature range in which planets are massive and cold enough to acquire and retain an atmosphere, while not too massive and cold to go runaway and explode into gas giants. While it is not perfect, it is seen in Figure \ref{fig:obs} that observed low-density super Earths seem to be restricted to the range predicted by our model. Some planets, on the other hand, seem to remain rocky, although they are in the Goldilocks region for acquiring and maintaining an atmosphere. A possible explanation to this diversity inside the Goldilocks region is giant impacts which remove gas \citep{InamdarHilke2015_new}.

An interesting outcome of the spontaneous shedding of the outer gas envelope after the nebula disperses is that atmospheres of planets just below the top line in Figure \ref{fig:obs} (which are on the verge of runaway accretion) are far from self gravitating. For our choice of $\gamma$, marginally self-gravitating atmospheres retain only $\approx 30\%$ of their mass after the nebula vanishes, as discussed in Section \ref{subsec:mass_loss}. However, this result is sensitive to our choice of $\gamma>4/3$, and for $\gamma<4/3$ we expect marginally self-gravitating atmospheres to retain most of their mass. Quantitaviely, for $\gamma=1.2$ \citep[as chosen by][]{LeeChiang2015}, atmospheres lose only $\approx 25\%$ of their mass (see Section \ref{subsec:mass_loss}) following disk dispersal.

Although \citet{LopezFortney2013} define a mass-flux curve qualitatively similar to our bottom line of Figure \ref{fig:obs}, with hot or small planets losing their atmosphere due to UV evaporation, the considerations leading to their curve are different from ours. In our model, we account for the cooling of the rocky core, and self-consistently couple the evolution prior to disk dispersal to the subsequent mass-loss.

\section{Conclusions and Discussion}\label{sec:conclusions}

In this work we analyzed the conditions in which rocky cores of a few $M_\Earth$ in mass acquire and retain voluminous atmospheres. Such atmospheres are necessary in order to explain some of the recently discovered {\it Kepler} low-density super Earths \citep[see, e.g.,][]{Lopez2012, Lissauer2013}.

We studied a scenario in which a pre-assembled rocky core accretes gas from the protoplanetary nebula. This scenario is relevant for the inner disk, where the assembly time for rocky cores is much shorter than the disk's lifetime \citep{GLS,Lee2014}. We also assumed that the gas density of the nebula is relatively low (MMSN), implying that atmosphere masses are determined by comparing their Kelvin-Helmholtz cooling time to the disk's dispersal time (the initial atmospheres which form adiabatically without cooling are negligible in mass). We found that in this scenario super Earths at $\sim 0.1$ AU orbits acquire atmospheres of a few \% in mass, in agreement with previous studies \citep{Lee2014,InamdarHilke,LeeChiang2015}. Despite the apparent consistency of this result with observations \citep[see, e.g.][]{Lopez2012}, subsequent evolution of the planets, following the nebula's dispersal, has to be considered as well.

We found that once the gas disk disperses, at time $t_{\rm disk}\sim 10\textrm{ Myr}$ \citep{Mamajek2009,WilliamsCieza2011,Alexander2014}, atmospheres shed their outer layers due to loss of pressure support from the disk, and due to their own cooling luminosity \citep[see also][]{OwenWu2015}. Consequently, super-Earth atmospheres shrink to a thickness comparable with the radius of the rocky core $R_{\rm rcb}\sim R_c$, on a timescale $\sim t_{\rm disk}$ (equivalently, the planets shrink to a radius $\approx 2 R_c$). When this {\it thin regime} is reached, we distinguish between two types of atmospheres: 
\begin{itemize}
\item {\it heavy envelopes}, with atmospheric masses (as a fraction of the core mass) $f\equiv M_{\rm atm}/M_c\gtrsim 5\%$, retain their mass and contract until they reach the maximum gas density after $\sim 1$ Gyr at most.   
\item {\it light envelopes}, on the other hand, with $f\lesssim 5\%$ are lost entirely due to the cooling of the underlying rocky core, which dominates the heat capacity (but can cool only once the atmosphere is thin). In this case, the mass-loss timescale is determined by the finite escape rate of molecules traveling at the speed of sound through the Bondi radius.
\end{itemize}
In addition to this spontaneous mass shedding, gas envelopes are also vulnerable to evaporation by high-energy stellar photons \citep[see, e.g.][]{Lopez2012, OwenWu2013}. In this work we examined both mass-loss possibilities and coupled them with the preceding nebular accretion phase.  

Our consistent treatment of accretion and evaporation allows us to relate the mass-loss timescale to the accretion time, and therefore to the disk's lifetime. Using these relations, we derived theoretical constraints on the planet's mass and equilibrium temperature (or equivalently, distance from the star) which enable the accretion and preserval of a significant atmosphere. Explicitly, we analytically identified a rather limited Goldilocks region in the temperature-mass plane, in which planets are massive and cold enough to obtain and retain an atmosphere, while not too massive or cold, so runaway gas accretion is avoided, and the planets do not become Jupiters. Observed low-density super Earths \citep[see, e.g.][]{WeissMarcy2014} are indeed concentrated in this theoretically allowed region, though some features of the observed super-Earth population are not explained by our model, and may be due to giant impacts \citep{InamdarHilke2015_new}.

It is noteworthy to mention \citet{Rogers2011}, who also coupled self-consistently core-nucleated accretion with the subsequent post-dispersal evolution, including mass loss. Despite the similar approach of both studies, we also point out the main differences. First, \citet{Rogers2011} focus on the scenario of formation beyond the snow line, at $T_{\rm eq}\approx100\textrm{ K}$, followed by migration to the current planet position of $T_{\rm eq}\sim 500-1000\textrm{ K}$. We, on the other hand, focus on in-situ formation at the current planet location. This difference affects both the accretion history and the composition of the rocky core, which is ice-rich in \citet{Rogers2011}. In addition, although UV-driven evaporation is treated similarly in both studies, this work also incorporates spontaneous mass loss, which is absent from \citet{Rogers2011}. Finally, we compare our model to planets which have reasonable constraints on their mass from transit time variations or radial velocity measurements, while \citet{Rogers2011} take a more general approach and use only the measured planet radius from {\it Kepler} to derive constraints on the possible mass and atmosphere fraction.

Our analytical model is simple, intuitive, and provides a consistent picture of gas accretion and evaporation, which seems to agree, at least approximately, with the observations. Nonetheless, there are several aspects of the model which deserve further attention:
\begin{itemize}
\item Although we coupled gas accretion and evaporation, we decoupled these processes from the assembly of rocky cores. At close orbits, planetesimal impacts may be ignored, as explained above, but giant impacts of protoplanets (or alternatively, inward migration of rocky cores) may be relevant, since isolation masses are small \citep[see, e.g.][]{GLS,InamdarHilke}.
\item 
We assumed a sharp decrease in the nebula's gas density, after which gas accretion terminates. The disk dispersal is more gradual, and planets may accrete gas from a depleted nebula \citep{LeeChiang3}. This caveat is briefly discussed in Section \ref{sec:transition}.
\item
This work focused on low-density gas nebulae (MMSN), while other works \citep{ChiangLaughlin2013,Lee2014,InamdarHilke} argue for higher gas densities (near the Toomre stability limit), based on high solid disk masses which were invoked to explain the {\it Kepler} observations. A better estimate of the initial gas density is thus required in order to constrain super-Earth formation scenarios (if the density is low enough though, it does not affect the results, as explained above).
Interestingly, our model can be used to test the presence (or absence) of high-density gas nebulae. In a dense nebula, the atmosphere mass is determined adiabatically, with cooling playing a minor role. By setting the runaway accretion condition $f\sim 1$ in Equation \eqref{eq:mass_adiabatic}, we find that the maximal stable core mass scales approximately as $M_c\propto T_{\rm eq}^{-1}$ in such nebulae (since cooling is unimportant, this result is robust and does not depend on the opacity or other details of our cooling model). In light nebulae, on the other hand, the atmosphere mass is determined by cooling, and the maximal stable core mass increases with temperature, as seen in Figure \ref{fig:obs} (top line). An improved set of observations, with accurate masses, and sampling a broad range of equilibrium temperatures (separations), can distinguish between the two models, and constrain the typical nebula density.
\item
The multi-dimensional flow around an accreting planet may alter the one-dimensional results presented here \citep{DangeloBodenheimer2013,Fung2015,Ormel2D,Ormel3D}. Specifically, \citet{Ormel3D} and \citet{Fung2015} find that in 3D simulations the atmosphere gas is replenished from the nebula on timescales shorter than the Kelvin-Helmholtz cooling time, resulting in lighter (i.e. higher entropy) envelopes. \cite{DangeloBodenheimer2013}, however, find that the envelope mass decreases only by a factor of roughly 2.
\end{itemize}
Addressing these points, together with a more realistic treatment of UV photo-evaporation \citep[see, e.g.,][]{OwenWu2013}, may improve our theoretical constraints on low-density super-Earth formation scenarios. In addition, a few extremely low density ``super-puffs''
\citep[see, e.g.,][and references within]{LeeChiang3} might require a different formation scenario.

\acknowledgements
This research was partially supported by ISF, ISA and iCore grants. We thank the TAPIR group at Caltech for warm hospitality during the initial stages of the research. We thank Eugene Chiang, Niraj Inamdar, Eve Lee, and the anonymous referee for valuable comments that improved the paper. 

\bibliography{bib}{}
\bibliographystyle{apj}

\end{document}